\documentclass[12pt]{article}  
\usepackage{graphicx}
\usepackage[figuresright]{rotating}
\usepackage{hyperref}  
\newcommand{\wwwspires}{http://www.slac.stanford.edu/spires/find/hep/www}  
\newcommand{\myhrefstring}{SPIRES}  
\newcommand{\beq}{\begin{equation}}  
\newcommand{\eeq}{\end{equation}}  
\newcommand{\be}{\begin{eqnarray}}  
\newcommand{\ee}{\end{eqnarray}}  
\newcommand{\Fdual}{\widetilde{F}}  
\newcommand{\fourint}[1]{\int\!\frac{d^4 #1}{(2\pi)^4}}  
\begin{document}  
\rightline{RUB-TPII-06/01}  
\rightline{hep-ph/0105173}  
\vspace{.3cm}  
\begin{center}  
\begin{large}  
{\bf Spin--dependent twist--4 matrix elements from the instanton \\  
vacuum: Flavor--singlet and nonsinglet} \\[1cm]  
\end{large}  
\vspace{1.0cm}  
{\bf Nam--Young Lee}$^{\rm a, 1}$,  
{\bf K.\ Goeke}$^{\rm a, 2}$  
{\bf and C. Weiss}$^{\rm b, 3}$  
\\[1.cm]  
$^{a}${\em Institut f\"ur Theoretische Physik II,  
Ruhr--Universit\"at Bochum, \\ D--44780 Bochum, Germany}  
\\
$^{b}${\em Institut f\"ur Theoretische Physik,
Universit\"at Regensburg, \\ D--93053 Regensburg, Germany}
\end{center}
\vspace{1.5cm}  
\begin{abstract}  
\noindent  
We estimate the twist--4 spin--1 nucleon matrix element $f_2$   
in an instanton--based description of the QCD vacuum. 
In addition to the flavor--nonsinglet we compute also the flavor--singlet   
matrix element, which appears in next--to--leading order of the  
$1/N_c$--expansion. The corresponding twist--3 spin--2 matrix elements $d_2$ 
are suppressed in the packing fraction of the instanton medium, 
$\bar\rho / \bar R \ll 1$. We use our results to estimate the leading $1/Q^2$
power corrections to the first moment of the proton and neutron
spin structure functions $G_1$, as well as the intrinsic charm contribution 
to the nucleon spin.   
\end{abstract}  
\vfill  
\rule{5cm}{.2mm} \\ 
{\footnotesize $^{\rm 1}$ E-mail: lee@tp2.ruhr-uni-bochum.de} \\  
{\footnotesize $^{\rm 2}$ E-mail: goeke@tp2.ruhr-uni-bochum.de} \\  
{\footnotesize $^{\rm 3}$ E-mail: weiss@tp2.ruhr-uni-bochum.de} \\  
\newpage  
\section{Introduction \label{introduction}}  
The structure of the nucleon as measured in deep--inelastic scattering  
is described by matrix elements of QCD operators of a certain  
twist (dimension minus spin). The moments of the structure functions 
at asymptotically large $Q^2$ are given by matrix elements of operators of  
twist 2, which have a simple interpretation as number densities of the  
quarks and antiquarks in the nucleon in the infinite--momentum frame. 
Power ($1/Q^2$--) corrections to the asymptotic result are determined by  
matrix elements of operators of twist 3 and 4. The latter describe either 
the correlation of the quark fields with the non-perturbative gluon  
field in the target, or quark--quark correlations. In the polarized 
case the twist--3 and 4 matrix elements of lowest spin are 
that of the twist--4 spin--1 operator 
\be  
\langle p, \lambda | \bar\psi_f \gamma_\alpha \Fdual^{\beta\alpha}  
\psi_f | p, \lambda \rangle &=& 2 M_N^2 \; f_{2,f} \; s^\beta ,  
\label{f2_def}  
\ee  
and of the twist--3 spin--2 operator 
\be  
\lefteqn{ \langle p , \lambda |   
\bar\psi_f  \left( \gamma^\alpha \Fdual^{\beta\gamma}    
+ \gamma^\beta \Fdual^{\alpha\gamma}  \right) \psi_f    
| p, \lambda \rangle - \mbox{traces}} && \nonumber \\  
&=& 2 \, d_{2, f} \; \left[ 2 p^\alpha p^\beta s^\gamma  
- p^\gamma p^\beta s^\alpha - p^\alpha p^\gamma s^\beta  
+ (\alpha \leftrightarrow \beta ) - \mbox{traces} \right] .  
\label{d2_def}  
\ee  
Here $\bar\psi_f, \psi_f$ denote the quark fields of flavor $f$, 
\be  
\Fdual^{\mu\nu} &\equiv& \frac{1}{2} \epsilon^{\mu\nu\rho\sigma}   
F_{\rho\sigma}  
\ee   
is the dual of the gauge field strength ($\epsilon^{0123} = 1$),  
$p$ the nucleon four--momentum ($p^2 = M_N^2$), $\lambda$ the
helicity, and $s = s(\lambda )$ the polarization  
vector of the nucleon state, which satisfies 
\beq  
s\cdot p \;\; = \;\; 0, \hspace{2cm}s^2 \;\; =\;\; -M_N^2 .  
\eeq  
The nucleon states are normalized according to 
$\langle p, \lambda |p', \lambda' \rangle =  
2p^0(2\pi)^3\delta^{(3)}({\bf p} -{\bf p}') \delta_{\lambda\lambda'}$. 
The matrix elements (\ref{f2_def}) and (\ref{d2_def}) can be taken either 
in a proton or neutron state.
\par
The matrix elements 
(\ref{f2_def}) and 
(\ref{d2_def}) together determine the leading power corrections to the  
first moment of the spin structure function $G_1$  
(the Bjorken and Ellis--Jaffe sum  
rules) \cite{Shuryak:1982pi,Ji:1994sv,Ehrnsperger:1994hh}. The twist--3 
matrix element $d_2$ also appears in the QCD expression for the third 
moment of the second spin structure function $G_2$, where it is not 
power--suppressed relative to the twist--2 part and can therefore be 
measured with 
good accuracy \cite{Bosted:2000tf}. By studying these matrix elements in 
theoretical models of the nucleon one may hope to get some insight 
into the transition from the region of asymptotically large $Q^2$  
($Q^2 \gg 1 \, {\rm GeV}^2$),  
where the $Q^2$--dependence of the structure functions is described   
by perturbative QCD, to the resonance region  
($Q^2 \approx 1 \, {\rm GeV}^2$) \cite{Edelmann:2000yp}. Eventually, this may  
help to understand also the discrepancy of the moments of the structure  
functions 
measured at $Q^2 \ge 1 \, {\rm GeV}^2$ with the GDH sum rule  
for photoabsorption ($Q^2 = 0$) \cite{Drechsel:1995az}.  
\par   
The twist--4 matrix element $f_2$, (\ref{f2_def}), also plays 
a role in the so-called ``intrinsic'' charm contribution 
to the nucleon spin \cite{Polyakov:1999rb}.  
Making use of an expansion in inverse powers of the charm quark  
mass, $1/m_c$, Franz {\it et al.}\ \cite{Franz:1999hw}, correcting 
an earlier results \cite{Halperin:1997jq,Araki:1998yk}, related the charm  
quark contribution to the axial current to the matrix element of 
the flavor--singlet twist--4 operator, (\ref{f2_def}). 
Model estimates of flavor--singlet $f_2$ can thus be used to  
estimate the intrinsic charm contribution to the nucleon spin. 
Knowledge of this contribution is 
a prerequisite for attempts to measure the gluon polarization  
through open charm production \cite{Baum:1996yv}.  
\par        
Here we report about an estimate of the spin--dependent  
twist--3 and 4  nucleon matrix elements, (\ref{f2_def}) and 
(\ref{d2_def}), within an instanton--based description of the QCD vacuum.   
In this approach the QCD ground state is approximately described as a  
``medium'' of topological vacuum fluctuations --- instantons and  
antiinstantons, with perturbative fluctuations about 
them \cite{Shuryak:1982ff,Diakonov:1984hh,Schafer:1998wv}.  
This picture explains the dynamical breaking  
of chiral symmetry \cite{Diakonov:1986eg},
which happens due to the fermionic zero modes associated with  
the individual (anti--) instantons, and manifests itself in the appearance of  
a dynamical mass of the quarks, accompanied by a coupling to Goldstone  
bosons --- the pions. Building on that, this picture explains   
a host of phenomenological data on ``vacuum structure''   
and hadronic correlation functions \cite{Schafer:1998wv}.  
It also gives rise to a successful description of the nucleon  
as a chiral soliton, in the formal limit of a large number of colors,   
$N_c \rightarrow \infty$ \cite{Diakonov:1988ty,Christov:1996vm}.  
\par
A crucial element in this approach is the fact that the instanton medium is 
dilute,  {\it i.e.}\ the average size of the (anti--) instantons, 
$\bar\rho \approx (600\, {\rm MeV})^{-1}$, is small 
compared to their average distance, $\bar R \approx (200 \, {\rm MeV})^{-1}$,
with $\bar\rho / \bar R \approx 1/3$. The existence of this small parameter
makes possible a systematic analysis of non-perturbative effects generated 
by the instantons.  
Matrix elements of quark--gluon operators of twist $> 2$ have been studied in 
this approach in Ref.\cite{Balla:1998hf}. In particular, it was shown 
there that the matrix element of the twist--3 operator, $d_2$, is 
suppressed relative to the twist--4 one, $f_2$, by a factor of 
$(\bar\rho / \bar R )^4$, and thus 
\beq  
\begin{array}{lcr}  
d_2 & \ll & f_2 . \\[1ex]  
\mbox{twist-3} & & \mbox{twist-4}  
\end{array}  
\eeq  
In Ref.\cite{Balla:1998hf} the value of $d_2$ was estimated to be of the  
order of $10^{-3}$; this prediction is confirmed by the recent results 
of the E155x experiment \cite{Bosted:2000tf}. Thus, in the instanton vacuum 
the parametrically leading higher--twist effects are due to the  
twist--4 matrix element $f_2$. In this letter we present a
quantitative estimate of $f_2$ in the instanton vacuum.
In addition to the flavor--nonsinglet 
nucleon matrix element, which appears in leading order of the $1/N_c$  
expansion and has been calculated in Ref.\cite{Balla:1998hf}, we 
calculate here also the flavor--singlet one. This allows us to  
make predictions for the power corrections to both the proton and
neutron spin structure functions $G_1$, as well as to estimate the intrinsic 
charm contribution to the nucleon spin. For a detailed description of the
foundations of the approach employed in the present investigation 
here we refer to Refs.\cite{Balla:1998hf,Diakonov:1996qy}.
\section{Twist--4 matrix elements from the instanton vacuum} 
{\it Chiral symmetry breaking by instantons.} 
Following Diakonov and Petrov \cite{Diakonov:1986eg,Diakonov:1986aj},  
the dynamical breaking of 
chiral symmetry in the instanton vacuum can be studied in the 
large--$N_c$ limit, where it manifests itself in the appearance 
of a dynamical quark mass, $M \neq 0$, accompanied by a  
coupling of the quarks to a pion (Goldstone boson) field, $\pi^a (x)$,  
$(a = 1, \ldots N_f^2 - 1)$ which arises
from the ``bosonization'' of the 't Hooft many--fermionic interaction 
induced by the instantons. The resulting low--energy dynamics is 
summarized by the effective Lagrangian 
\beq  
L_{\rm eff} \;\; = \;\; \bar\psi (x) \left[ i\!{\not\partial}  
- M F(\partial^2) \, U^{\gamma_5}(x)
F(\partial^2 ) \right] \psi(x),  
\label{L_eff} 
\eeq  
where 
\beq
U^{\gamma_5}(x) \;\; = \;\; \frac{1 + \gamma_5}{2} U(x)
+ \frac{1 - \gamma_5}{2} U^\dagger (x), 
\hspace{4em} U(x) \;\; = \;\; e^{i \pi^a (x) \tau^a} , 
\eeq
is a unitary matrix variable containing the pion field,
and $F(p^2 )$ are form factors, proportional to the  
Fourier transform of the zero mode of the (anti--) instantons,  
which drop to zero for spacelike momenta of the order 
of $-p^2 \sim \bar\rho^{-2}$. It is crucial that the 
dynamical quark mass, $M$, is parametrically small compared to the 
ultraviolet cutoff, $\bar\rho^{-1}$: 
\beq  
M\bar\rho\; \sim \;\left(\frac{\bar\rho}{\bar R}\right)^2 . 
\eeq  
In particular, this implies that for spacelike quark momenta of the 
order $M \ll \bar\rho^{-1}  \approx 600\, {\rm MeV}$  
one may neglect the form factors 
and consider the quark--pion coupling as effectively pointlike. 
\par 
{\it Nucleon as chiral soliton.} 
The effective Lagrangian (\ref{L_eff}) serves as a basis for the 
calculation of correlation functions in the instanton vacuum 
within the $1/N_c$--expansion \cite{Diakonov:1986eg,Diakonov:1986aj}. 
In particular, it gives rise to a picture of the nucleon as a chiral 
soliton \cite{Diakonov:1988ty}. Baryon correlation functions
in the large--$N_c$ limit are characterized by a classical pion field, 
which in the baryon rest frame is of ``hedgehog'' form 
\beq 
U_c ({\bf x}) \;\;=\;\; e^{i n^a \tau^a P(r)} ,  
\hspace{4em} r \; \equiv \; |{\bf x}|,  
\hspace{4em} {\bf n} \; = \; \frac{\bf x}{r} . 
\label{hedge} 
\eeq 
Baryon states of definite spin/isospin and momentum quantum numbers emerge 
from quantizing the collective rotations and translations of the 
classical soliton 
\beq 
U({\bf x}, t) \;\; = \;\;  
R (t) U_c ({\bf x}-{\bf X}(t)) R^\dagger (t) . 
\label{U_time_dependent}
\eeq 
Since the soliton moment of inertia and mass are of order $N_c$, 
the angular velocity of the collective rotation, 
\beq  
\Omega\;\; = \;\; -iR^\dag\dot{R} ,  
\label{angular_vel} 
\eeq
as well as the linear velocity of the translation, are of order 
$1/N_c$, the quantization of the collective motion can be performed
within a $1/N_c$ expansion \cite{Diakonov:1988ty}. 
The $\Omega^2$--contribution to the baryon mass gives rise to the 
$N$--$\Delta$ mass splitting, which is of order $1/N_c$.
This approach allows also to compute matrix elements of
operators between baryon states. Technically, this is done by
representing the matrix element as a functional integral over
the collective translations/rotations, which is then evaluated
by expanding the integrand in $\Omega$. For details we refer to
Refs.\cite{Diakonov:1988ty,Christov:1996vm}.  
%
%
\begin{figure}[t]
\begin{center}
\includegraphics[width=12.0cm,height=1.86cm]{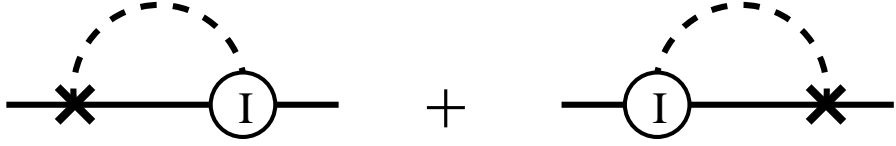}
\end{center}
\caption[]{Schematic illustration of the contributions to the matrix element 
of the quark--gluon operator, (\ref{f2_def}), in the instanton vacuum. 
The cross denotes the quark--gluon operator, the solid line the quark 
propagator, the dashed line the gauge field of the (anti--) instanton, 
and the circle the chirality--flipping 't Hooft interaction of the 
instanton with the quark mediated by the zero modes. The picture 
shows the situation for one light quark flavor; for $N_f$ flavors the
't Hooft interaction is a $2 N_f$--fermionic vertex, which
at large $N_c$ is equivalent to a coupling of the quark to the pion field
(``bosonization''). Using the equations of motion of the effective low--energy
theory, the instanton--induced vertex shown here can be converted to 
the chirally even effective operator (\ref{f2_equiv}).
For details, see Refs.\cite{Balla:1998hf,Dressler:1999zi}.}
\label{fig_effop}
\end{figure}
\par 
{\it Twist--4 quark--gluon operator in the instanton vacuum.} 
The instanton vacuum, being 
a microscopic model of the non-perturbative fluctuations of the gluon  
field, allows to estimate hadronic matrix elements of quark--gluon operators, 
such as the twist--4 operator of Eq.(\ref{f2_def})  
\cite{Diakonov:1996qy,Balla:1998hf}. It is understood that 
the QCD operators are normalized at the scale defined by the
inverse average instanton size, $\mu = \bar\rho^{-1}$.
The main idea of this approach is that because of the diluteness of the 
instanton medium the gauge field operator in (\ref{f2_def}) can  
effectively be replaced by the field of a {\it single} 
(anti--) instanton, which then interacts with the quark fields through 
its zero modes (this is schematically illustrated in Fig.~\ref{fig_effop}).
As a result, the original QCD quark--gluon operator is replaced by a 
chirality--flipping ``effective operator'', which is to be evaluated in 
the effective theory where chiral symmetry is spontaneously 
broken. As has been discussed in Refs.\cite{Balla:1998hf,Dressler:1999zi}, 
this effective operator can be brought to a chirality--conserving
form by making use of the equations of motion of the effective theory.
For the QCD operator of interest, 
$\bar\psi_f \gamma_\alpha \Fdual_{\beta\alpha} \psi_f$,
the resulting operator is of the form \cite{Balla:1998hf}
\beq  
- f_2^{\rm quark} \;   
\bar\psi_f \gamma_\beta \gamma_5 \partial^2 \psi_f .
\label{f2_equiv}  
\eeq  
It is understood here that the momenta of the quark fields are
of the order $p \sim M \ll \bar\rho^{-1}$. Here $f_2^{\rm quark}$ 
denotes the matrix element of the
original chirality--flipping effective operator in a quark state, 
given by
\beq  
f_2^{\rm quark} \;\; = \;\; I_1 (-p^2 ) |_{p^2 = M^2} ,  
\label{f2_from_I}  
\eeq
where $I_1 (-p^2)$ denotes the quark loop integral defined
by ($\bar p$ and $\bar k$ are Euclidean momenta, with $\bar p^2 = - p^2$)
\beq  
I_1 (\bar p^2 ) \;\; = \;\; \bar\rho^2 \fourint{\bar k}   
\frac{F(\bar p - \bar k) {\cal G} (\bar k)}{(\bar p - \bar k)^2}   
\left[ -\frac{3}{4} \frac{|\bar k|}{|\bar p|} C_1^{(1)} (\cos\theta )   
+ \frac{1}{2} C_2^{(1)} (\cos\theta ) \right] ,  
\label{f2_integral}  
\eeq  
in which $C_n^{(1)}$ are the Gegenbauer polynomials (4--dimensional  
spherical harmonics) of argument 
$\cos\theta = (\bar k\cdot \bar p)/(|\bar k||\bar p|)$, and the function 
${\cal G} (\bar k)$ 
describes the Fourier transform of the (dual) field strength of 
a single (anti--) instanton in singular gauge (see Ref.\cite{Balla:1998hf} 
for details): 
\be 
{\cal G} (\bar k) 
&=& 32 \pi^2 \left[ \left(\frac{1}{2} + \frac{4}{t^2} \right)  
K_0 (t) + \left( \frac{2}{t} + \frac{8}{t^3} \right) K_1 (t)  
- \frac{8}{t^4} \right], \hspace{1cm} t \; = \; |\bar k|\rho ,  
\label{G}  
\ee 
with $K_n (t)$ the modified Bessel functions of the second kind. The  
numerical value of $f_2^{\rm quark}$ in the limit $M\bar\rho \rightarrow 0$ 
is \cite{Balla:1998hf}
\beq  
f_2^{\rm quark} \;\; = \;\; \frac{3}{5} \;\; = \;\; 0.6 .  
\label{f2_quark}  
\eeq
\par
The instanton--induced effective operator (\ref{f2_equiv}) has the
form of the axial current operator, but with an additional contracted
derivative acting on the quark fields. Note that the coefficient,
$f_2^{\rm quark}$, is of order unity in the packing  
fraction of the instanton medium, $\bar\rho / \bar R$. 
Technically speaking, this
happens because the quark loop integral $I_1$, (\ref{f2_integral}),
contains a would--be quadratic divergence, which is regularized
by the instanton form factors and cancels the overall factor 
$\bar\rho^2$ in front of the integral. This implies that the
hadronic matrix elements of this operator will also be parametrically
large in the instanton packing fraction.
We remark that for the twist--3 operator in the matrix element, 
$d_2$, (\ref{d2_def}), the effective operator 
would be proportional to $(M\bar\rho )^2 \sim (\bar\rho / \bar R)^4$ and 
thus parametrically suppressed. 
\par 
{\it Nucleon matrix elements: $1/N_c$ expansion.}
The nucleon matrix elements of the effective operator, (\ref{f2_equiv}),
can be computed using the same techniques as have been used in 
the calculation of nucleon matrix elements of vector and axial vector
current operators (for a review, see \cite{Christov:1996vm}).  
In particular, we shall exploit the close relation of the
effective operator, (\ref{f2_equiv}), with the axial current operator.
It is known that within the chiral soliton picture of the nucleon
the isovector and isoscalar component of the nucleon axial coupling
constant appear in different orders in the expansion in powers of the
angular velocity of the collective rotation of the soliton, 
$\Omega \sim 1/N_c$, (\ref{angular_vel}). We therefore need to
consider separately the isovector and isoscalar components of the
twist--4 nucleon matrix element $f_2$, (\ref{f2_def}),
\beq
f_2^{(3)} \;\; = \;\; f_{2, u} - f_{2, d}, 
\hspace{2cm}
f_2^{(0)} \;\; = \;\; f_{2, u} + f_{2, d} 
\label{f2_vec_scal}
\eeq
(the matrix elements on the R.H.S.\ refer to the proton). We restrict
ourselves for the moment to the $SU(2)$ flavor group ($u, d$), the 
extension to $SU(3)$ will be discussed below.
For the isospin components defined by (\ref{f2_vec_scal}) the $N_c$--counting 
can immediately be inferred from that of the corresponding
axial coupling constants of the nucleon: 
\beq 
M_N^2 f_2^{(3)} \;\; \sim \;\; N_c,  
\hspace{2cm} 
M_N^2 f_2^{(0)} \;\; \sim \;\; 1. 
\eeq 
The isovector component of the spin--dependent twist--4 matrix element
is leading in the $1/N_c$ expansion.
\par 
{\it Isovector matrix element} $f_2^{(3)}$. 
We now make a quantitative estimate of the isovector twist--4 nucleon matrix 
element, $f_2^{(3)}$. This can be done by considering the 
hypothetical limit of large soliton size, in which one can perform an 
expansion of the matrix element in gradients of the classical pion 
field, (\ref{hedge}). The result for $f_2^{(3)}$ in this approximation 
can immediately be written down by analogy with the well--known result 
for the isovector axial coupling, $g_A^{(3)}$ \cite{Diakonov:1988ty},
\be  
g_A^{(3)} &=& \frac{4 N_c M^2 J}{9} \int d^3 x \;   
{\rm tr}\,\left[ -i \tau^a U_c^\dagger ({\bf x}) 
\partial_a U_c ({\bf x}) \right] ,  
\label{ga_grad}  
\ee   
where $J$ denotes the (Euclidean) quark loop integral   
\beq  
J \;\; = \;\; \fourint{\bar k} \;    
\frac{ F^2 (\bar k) \left[ F^2 (\bar k) -  \bar k^2 F(\bar k) F' (\bar k) 
\right]}  
{\left[\bar k^2 + M^2 F^4 (\bar k) \right]^2} .  
\label{J} 
\eeq  
(Note that the PCAC relation relates this integral to the pion decay 
constant, $4 N_c M^2 J = F_\pi^2$.) The corresponding result for the
isovector matrix element of the operator (\ref{f2_equiv}) is 
\be  
M_N^2 f_2^{(3)} &=& \frac{4 N_c M^2 f_2^{\rm quark} J_1}{9} \int d^3 x \;   
{\rm tr}\,\left[ -i \tau^a U_c^\dagger ({\bf x}) 
\partial_a U_c ({\bf x}) \right] ,  
\label{f2_grad}  
\ee   
where 
\be  
J_1 &=& \fourint{\bar k} \;    
\frac{ \bar k^2 F^2 (\bar k) \left[ F^2 (\bar k) -  \bar k^2 
F (\bar k) F' (\bar k) \right]}  
{\left[\bar k^2 + M^2 F^4 (\bar k) \right]^2} ,  
\label{I_f_2}  
\ee  
which differs from $J$, (\ref{J}), by an additional power of the 
quark virtuality, $\bar k^2$, in the integrand. 
For the ratio of $f_2^{(3)}$ to $g_A^{(3)}$ we thus obtain  
\beq  
\frac{M_N^2 f_2^{(3)}}{g_A^{(3)}} \;\; = \;\; 
-\frac{f_2^{\rm quark} J_1}{J} ,  
\label{f2_grad_as_ratio}  
\eeq  
With the standard parameters of the instanton vacuum, 
$\bar\rho = (600\, {\rm MeV})^{-1}$ and $M = 350\, {\rm MeV}$,
we find
\beq  
\frac{f_2^{\rm quark} J_1}{J} \;\; =\;\; 0.49 \, \bar\rho^{-2} .  
\eeq  
With the experimental values $M_N = 940\, {\rm MeV}$ and   
$g_A^{I=1} = 1.25$, Eq.(\ref{f2_grad_as_ratio}) thus   
gives\footnote{In the numerical estimate quoted in Ref.\cite{Balla:1998hf},
$f_2^{(3)} = -0.11$, some contributions to the gradient expansion of the 
matrix element were missed. Hence the differences in the numerical values.}  
\beq  
f_2^{(3)} \;\; = \;\; -0.25 \; .  
\label{f2_3_num} 
\eeq
\par
The integral $J_1$ in the gradient expansion of $f_2^{(3)}$, (\ref{I_f_2}),
contains a would--be quadratic divergence, {\it i.e.}, it is parametrically
of the order $J_1 \sim \bar\rho^{-2}$, contrary to the integral $J$
in $g_A^{(3)}$, which is depends only logarithmically 
on $\bar\rho^{-1}$. Thus, in $J_1$ the dominant contribution comes
from quark momenta of the order of the UV cutoff, $\bar k \sim \bar\rho^{-1}$.
In principle the representation of the instanton--induced effective
operator by the local operator, (\ref{f2_equiv}), is accurate only
for quark momenta of the order $\bar k \sim M \ll \bar\rho^{-1}$; 
for momenta of the order $\bar\rho^{-1}$ one should take into account the
momentum--dependence (non-locality) of the instanton--induced
effective operator. One should keep in mind that, anyway, the precise form
of the UV cutoff of the effective low--energy theory depends on 
the details of the approximations made in the description of the instanton 
medium; only the gross features can be assumed to be generic. Thus, 
results for quantities given by ``quadratically divergent'' integrals, such 
as $f_2^{(3)}$, should generally be regarded as rough estimates ($\pm 50 \%$). 
Whether or not to include the form factors in the effective operator is just 
part of this larger uncertainty. We have verified that using instead of 
(\ref{f2_equiv})
the ``exact'' non-local operator $-\bar\psi \gamma_\beta \gamma_5    
\left[ I_1 (\partial^2 ) / F(\partial^2 ) \right] \partial^2 \psi$
changes the numerical result for $f_2^{(3)}$ by less than 10\%, so 
using the local approximation to the operator seems completely justified.
\par 
{\it Isoscalar matrix element} $f_2^{(0)}$. 
The calculation of the $1/N_c$--suppressed isoscalar component of the  
twist--4 matrix element, $f_2^{(0)}$, is somewhat more involved, since  
it requires to take into account the time dependence of the 
saddle--point pion field, (\ref{U_time_dependent}), to first order in the 
angular velocity of the soliton, $\Omega$, (\ref{angular_vel}).
An important question is the degree of the would--be ultraviolet 
divergence of the isoscalar nucleon matrix element, $f_2^{(0)}$. This question
can be investigated by gradient expansion. Expanding the average over
quark fields of the flavor--singlet version of the operator 
(\ref{f2_equiv}),
\beq
(-i) f_2^{\rm quark}
N_c \, {\rm Sp} \left[ \, 1_{\rm Flavor}\; \gamma_\beta \gamma_5 \partial^2
\frac{1}{i\!{\not\partial} - M F(\partial^2 ) U^{\gamma_5} F(\partial^2 )}
\right] ,
\label{singlet_trace}
\eeq
in gradients of the (space-- and time--dependent) background pion field, 
one finds that the functional trace is at most {\it logarithmically}
divergent. This implies that the matrix element, which is obtained
by substituting in (\ref{singlet_trace}) the slowly rotating pion
field (\ref{U_time_dependent}), and expanding to first order in the angular
velocity, $\Omega$, can also be at most logarithmically divergent.
This circumstance is very fortunate, as it allows us to neglect the 
instanton--induced form factors in the quark--pion coupling 
in the Lagrangian, (\ref{L_eff}), and simply apply an external UV 
regularization in the form of a Pauli--Villars subtraction.
(If the matrix element were quadratically divergent, as $f_2^{(3)}$ is,
the result would have been strongly dependent on the specific form of 
ultraviolet cutoff applied.) With this simplification the calculation of 
the matrix element of the instanton--induced effective operator 
(\ref{f2_equiv}) becomes completely analogous to that of the isovector 
axial coupling constant, $g_A^{(0)}$ \cite{Christov:1994ny,Blotz:1996wi}, 
and can be performed using the same techniques.
\par
A new feature in the calculation of the matrix element 
of the instanton--induced operator (\ref{f2_equiv}) compared to 
that of the axial current, $\bar\psi\gamma_\beta\gamma_5\psi$,
is that with (\ref{f2_equiv}) contributions of order $\Omega^1$ can 
appear from the action of the derivatives contained in the operator 
on the time--dependent  isospin rotation matrices, $R(t)$. Such contributions  
were discussed in Ref.\cite{Pobylitsa:1999tk} in the context of the   
calculation of the isovector unpolarized quark distribution in the  
nucleon, and we refer to this article (in particular Section 3.1) for a  
detailed description of the $\Omega$-- ($1/N_c$--) expansion. 
After performing the $\Omega$--expansion one expresses the 
remaining Green functions in the static background pion field, 
(\ref{hedge}), in a basis of eigenfunctions of the quark single--particle 
Hamiltonian in the:
\be 
H &=& -i \gamma^0 \gamma^i \nabla_i + \gamma^0 M U^{\gamma_5} ,
\\ 
H |n\rangle &=& E_n |n\rangle . 
\ee 
The result for the isoscalar matrix element then reads
\be  
f_2^{(0)} &=& \frac{N_c \, f_2^{\rm quark}}{M_N^2} 
\left\{ -\frac{1}{4I}  
\sum_{\scriptstyle n \atop \scriptstyle {\rm non-occup.}} 
\sum_{\scriptstyle m \atop \scriptstyle {\rm occup.}} 
\frac{1}{E_n-E_m}  
\right. 
\nonumber \\ 
&& \phantom{\frac{N_c \, f_2^{\rm quark}}{M_N^2}} 
\;\;\;\;\; \times \left[  
\langle n|\sigma_3(E_m^2- {\bf p}^2 )|m\rangle\langle m|\tau_3 |n \rangle   
+\langle n|\tau_3|m\rangle \langle m|\sigma_3(E_n^2- {\bf p}^2 )|n\rangle  
\right] 
\nonumber \\ 
&& \phantom{\frac{N_c \, f_2^{\rm quark}}{M_N^2}} 
\left. 
-\frac{1}{2 I} 
\sum_{\scriptstyle n \atop \scriptstyle {\rm occup.}} 
E_n \langle n|\sigma_3\tau_3|n\rangle \right\} . 
\label{sum} 
\ee  
Here $I$ is the moment of inertia of the classical soliton,  
see Ref.\cite{Diakonov:1988ty}. 
The first term in the braces is a double sum over quark levels; it 
coincides with the expression for $g_A^{(0)}$ up to the insertions 
of the operator $E^2 - {\bf p}^2$ in the single--particle matrix elements, 
which come from the contracted derivative in the operator (\ref{f2_equiv}). 
The second term represents the abovementioned ``new'' contribution 
which results from the action of the derivative in the operator on the 
rotational matrices in the collective quantization. Up to the factor  
$E_n$ inside the single sum over quark levels this contribution is 
identical to the expression for the {\it isovector} axial coupling 
constant, $g_A^{(3)}$.
\par 
In Eq.(\ref{sum}) the first term (the double sum over quark levels)
turns out to be UV--finite and does not require regularization.
The second term (the simple sum over levels, related to the isovector
axial coupling constant), as it stands, contains a linear divergence
when regularized {\it e.g.}\ with an energy cutoff. At the same time
one observes an anomaly--type phenomenon in the sense that the sum over
occupied quark levels is not equal to minus the sum over non-occupied ones,
as it should be on grounds of the analyticity properties of the 
single--particle Green functions. However, both the linear divergence
and the ``anomaly'' disappear when performing a Pauli--Villars (PV) 
subtraction, leaving behind a finite sum with usual behavior
when summing over non-occupied instead of occupied states, in agreement
with the above statement that the matrix element is at most logarithmically
divergent. A similar phenomenon was encountered in the calculation of 
matrix elements of twist--2 operators of spin $>2$ in the chiral 
quark--soliton model in Ref.\cite{Diakonov:1997vc}. 
In that case, too, the superficial 
power divergences and the ``anomalies'' simultaneously disappeared 
after PV subtraction. Note that the PV subtraction, contrary to 
regularization methods based on an energy cutoff, does not spoil
the analyticity properties of the regularized sums.
\par 
The remaining PV regularized sums can be performed numerically, 
using the Kahana--Ripka method of diagonalizing the single--particle 
Hamiltonian in a basis of free quark states.  
The value of the PV cutoff
is determined by fitting the pion decay constant \cite{Diakonov:1997vc},  
$M_{PV} = 557\, {\rm MeV}$. For the numerical estimate we use 
the variational soliton profile of 
Refs.\cite{Diakonov:1988ty,Diakonov:1997vc}. 
We obtain a numerical value of the flavor--singlet twist--4  
matrix element of 
\beq  
f_2^{(0)} \;\; = \;\; 0.01 . 
\label{f2_0_num} 
\eeq  
The flavor--singlet matrix element is more than an order of magnitude  
smaller than the flavor--nonsinglet one, (\ref{f2_3_num}), which reflects 
the fact that the former is given by a ``logarithmically divergent'', the 
latter by a ``quadratically divergent'' expression. 
\par
{\it Comparison with results of other approaches.} The instanton result 
for the isovector twist--4 matrix element, $f_2^{(3)}$, agrees 
both in sign and in magnitude with the QCD sum 
rule results of Refs.\cite{Balitsky:1990jb,Stein:1995si}. 
However, we disagree with these authors on the sign of the isoscalar
matrix element (for a critical discussion of the QCD sum rule
calculations, see Refs.\cite{Ioffe:1997ey}).
We also disagree in the sign of the isovector matrix element
with the bag model estimate of Ref.\cite{Ji:1994sv};
however, this model can hardly claim to give a realistic description
of the quark--gluon correlations giving rise to the twist--4 matrix
elements; {\it e.g}, it does not respect the QCD equations of motion.
\par  
{\it Generalization to the $SU(3)$ flavor group.} 
So far we have calculated the twist--4 matrix elements assuming 
the $SU(2)$ flavor group ($u, d$--quarks only). The results 
can easily be generalized to the $SU(3)$ case (including also $s$ quarks). 
Here we consider the simplest scenario of perfect $SU(3)$ symmetry, 
{\it i.e.}, zero strange quark mass.
In the chiral quark--soliton model the leading--$N_c$ calculation 
gives a simple relation between the $SU(3)$ triplet and
octet couplings, as the differences between them appear only due
to the different states of collective rotations of the soliton.
The ratio of the triplet and the octet part of the nucelon matrix element
$f_2$ is therefore given by the ratio of the matrix elements of 
the respective Wigner $D$--functions in the rotational state corresponding
to the nucleon spin, isospin and the hypercharge, 
$|\mbox{rot}\rangle = |S=T=1/2, S_3 = 1/2, T_3 = 1/2 \rangle$ 
(for details see \cite{Christov:1996vm}): 
\begin{eqnarray}  
\frac{f_2^{(8)}}{f_2^{(3)}}&=&
\frac{\langle\, \mbox{rot} \,
|\,D^8_{83}\,|\,\mbox{rot}\,\rangle}{\langle
  \,\mbox{rot}|\,D^0_{33}\,|\,\mbox{rot}\rangle}
\;\; = \;\; \frac{3}{7}.
\end{eqnarray}
From these relations we obtain the following numerical results 
for the twist--4 matrix elements in the $SU(3)$ symmetric case: 
\be  
f_2^{(0)} |_{SU(3)} &=& \phantom{-} 0.01   
\label{f20_su3} \\   
f_2^{(3)} |_{SU(3)} &=& -0.25
\label{f23_su3} \\  
f_2^{(8)} |_{SU(3)} &=& -0.11.
\label{f28_su3}
\ee  
\section{Power corrections to proton and neutron spin structure functions}
With the results obtained in the previous section we can now
estimate the $1/Q^2$--corrections to the first moments of the
proton and neutron spin structure functions, $G_1^{\rm p, n}$.
The expressions obtained from the operator product expansion
of QCD at tree level to order $1/Q^2$ 
are \cite{Shuryak:1982pi} \footnote{Note that there 
is a mistake in Ref.\cite{Shuryak:1982pi} concerning the coefficients in 
this formula, as was noted in \cite{Ji:1994sv}; see also   
Ref.\cite{Ehrnsperger:1994hh}.} 
\beq
\int_0^1 dx \, G_1^{p(n)} (x, Q^2 )
=\frac{1}{2}a_0^{p(n)}+\frac{M_N^2}{9Q^2}
\left(a_2^{p(n)}+4d_2^{p(n)}+4f_2^{p(n)}\right) .
\label{g1_pn}
\eeq
Inclusion of radiative corrections would lead to a logarithmic 
$Q^2$--dependence of the coefficients. Here $a^{(0)}_n$ 
denote the spin--dependent matrix elements of the twist--2 
spin--$(n + 1)$ axial vector operator; the proton and neutron
matrix elements (which include the quark charges) are
given in terms of the $SU(3)$ singlet, triplet and octet 
components as
\begin{equation}
a_0^{p(n)} = \pm\frac{1}{6}a^{(3)}+\frac{1}{18}a^{(8)}+\frac{2}{9}a^{(0)},
\label{a0_pn}
\end{equation}
with $+(-)$ for proton (neutron). We quote the expressions
for three light quark flavors ($u, d, s$).
For $n = 0$ the $a_n$ coincide with the 
axial coupling constants of the nucleon:
\beq  
a_0^{(3)} \;\; = \;\; g_A^{(3)}, \hspace{4em}  
a_0^{(8)} \;\; = \;\; g_A^{(8)}, \hspace{4em}  
a_0^{(0)} \;\; = \;\; g_A^{(0)}.  
\eeq  
Of the power corrections the term proportional to $a_2^{p(n)}$
are the target mass corrections, while the terms proportional
to $d_2^{p(n)}$ and $f_2^{p(n)}$ represent the dynamical 
higher--twist corrections; the proton and neutron matrix elements
are defined in analogy to Eq.(\ref{a0_pn}). 
\par
\begin{figure}[ht]  
\begin{center}  
\begin{tabular}{c}
\includegraphics[width=12cm,height=8.4cm]{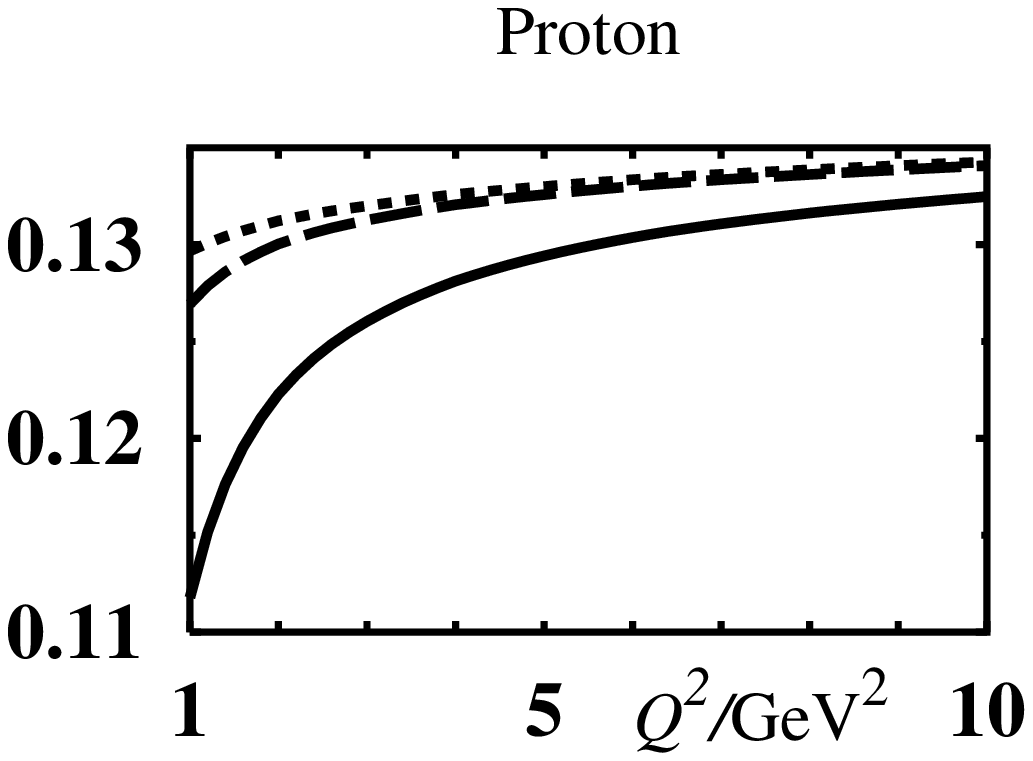}
\\
\includegraphics[width=12cm,height=8.4cm]{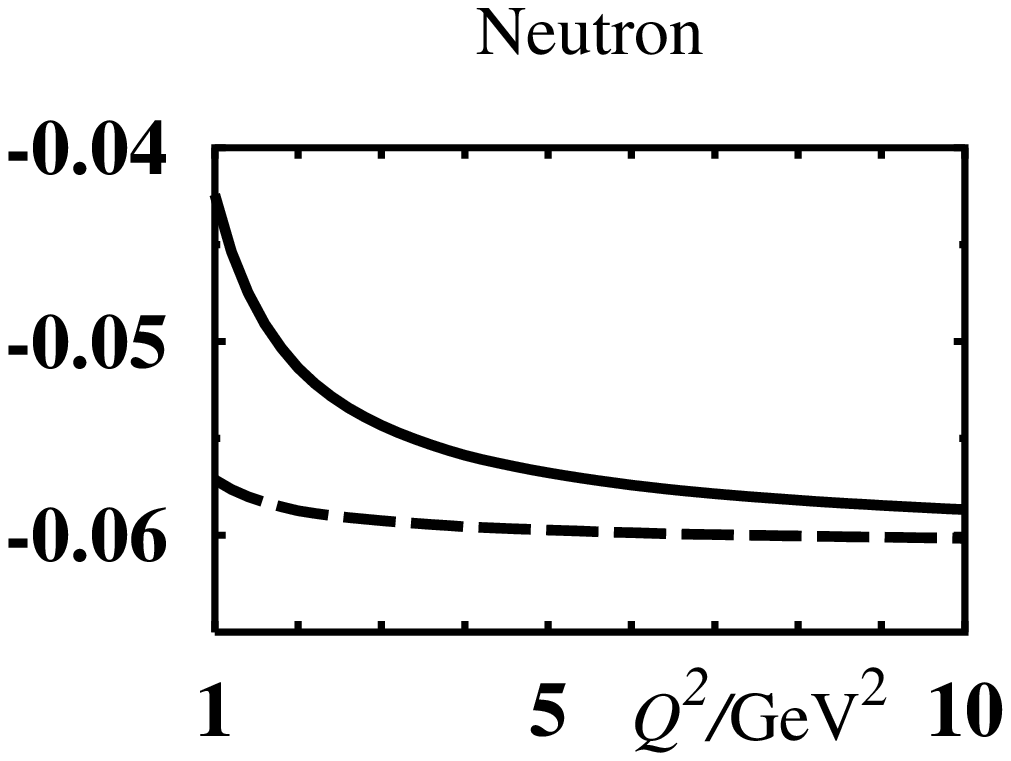}
\end{tabular}
\end{center}  
\caption{The $Q^2$--dependence of the first moment of the polarized structure 
function $G_1$  
for the proton (top) and neutron (bottom). {\it Dashed lines:}
Twist--2 contribution according to the GRSV 2000 NLO parameterization
(standard scenario) \cite{Gluck:2000dy}. {\it Dotted lines:} Sum of twist--2
contribution and target mass corrections. {\it Solid lines:} 
Total result, including also the twist--4 contribution due to
the matrix elements $f_2$, as estimated in the instanton vacuum.
(In the case of the neutron the target mass corrections to the 
first moment are very small, so we show only the pure twist--2 
contribution.)}
\label{fig_p_n}  
\end{figure}

We now evaluate the dynamical power corrections using the
instanton vacuum results.
Following the basic philosophy of the instanton vacuum we put the
parametrically suppressed twist--3 spin--2 matrix 
elements to zero (these matrix elements were estimated in
Ref.\cite{Balla:1998hf} 
to be of the order of $\sim 1 \%$ of the twist--4 ones).
For the twist--4 matrix elements $f_2^{p(n)}$ we use the
results (\ref{f20_su3})--(\ref{f28_su3}), which imply
\beq
f_2^p \;\; = \;\; -0.046 , 
\hspace{4em}
f_2^n \;\; = \;\; 0.038 .
\eeq
To evaluate the twist 2 contribution we use the GRSV 2000 
parameterization \cite{Gluck:2000dy} (``standard scenario''), 
which includes the radiative corrections to the coefficient
functions in NLO.\footnote{To evaluate the target mass corrections
we use instead of the twist--2 matrix element $a_2^{p(n)}$ the third 
moment of the twist--2 part of $G_1^{p(n)}$ calculated from the GRSV 2000
parametrization; the difference between the two is irrelevant in the 
present context.} 
The results are shown in Fig.\ref{fig_p_n}. One sees that the dynamical 
twist--4 contribution
has a rather small effect on the structure functions down to 
$Q^2$ of the order of $1\, {\rm GeV}^2$, in particular in the neutron.
Note that the quantitative details may change when $SU(3)$ symmetry
breaking effects are included in the twist--4 contribution.
%
%
\clearpage
\par
\begin{figure}[ht]  
\begin{center}  
\includegraphics[width=12cm,height=8.4cm]{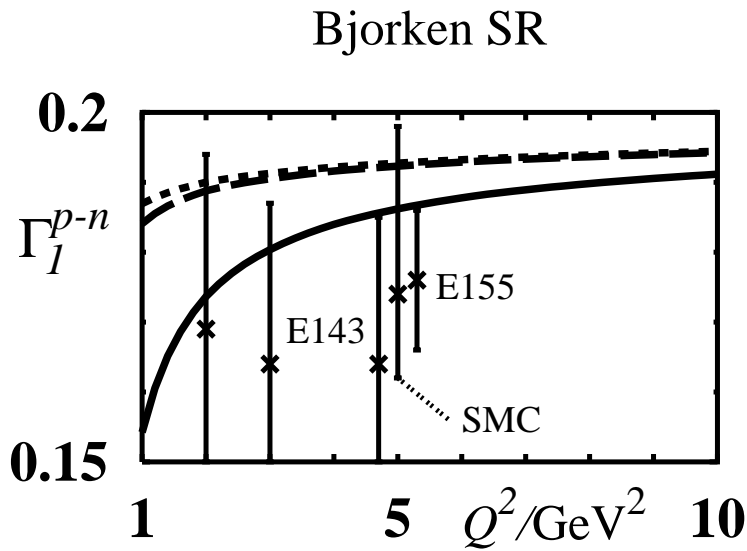}
\end{center}  
\caption{Same as Fig.\ref{fig_p_n}, but for the difference of proton
and neutron structure functions (Bjorken sum rule). Also shown are the
experimental results quoted by the E143 collaboration 
at $Q^2 = 2, 3$ and $5 \, {\rm GeV}^2$ \cite{Abe:1998wq}, and 
by the SMC \cite{Adeva:1998vv} and E155 \cite{Anthony:2000fn}
collaborations at $Q^2 = 5 \, {\rm GeV}^2$. (The abscissae of the 
data points at $Q^2 = 5\, {\rm GeV}^2$ have been shifted slightly
in order to separate them in the plot.)}
\label{fig_bjorken}  
\end{figure}

In Fig.\ref{fig_bjorken} we show the result for the first moment
of the difference of proton and neutron structure functions,
$G_1^p - G_1^n$ (the Bjorken sum). This quantity is of particular 
interest, since the Bjorken sum rule is a rigorous prediction of QCD,
and the radiative corrections have been calculated to order $\alpha_s^3$
\cite{Larin:1991tj}. We again use the GRSV2000 NLO parametrization
to estimate the twist--2 part and target mass corrections. 
The power corrections now receive contributions only from the 
flavor--triplet twist--4 matrix element, $f_2^{(3)}$.
Also shown in Fig.\ref{fig_bjorken} are the experimental results
obtained in the analyses of the SMC \cite{Adeva:1998vv}, 
E143 \cite{Abe:1998wq}, and E155 \cite{Anthony:2000fn} experiments.
One observes that the relatively large twist--4 correction obtained
from the instanton vacuum improves the agreement of the theoretical
prediction with the data; however, the present experimental errors
are too large to allow for definite conclusions.
%
%
\section{Intrinsic charm contribution to the proton spin}  
The isoscalar component of the twist--4 matrix element, $f_2^{(0)}$,
(\ref{f2_def}), plays a role in the nucleon spin structure functions
aside from power ($1/Q^2$--) corrections, namely in determining the 
intrinsic charm contribution to the nucleon spin.
Up to power corrections, the first moment of the proton 
spin structure function for three light flavors ($u, d, s$)
is given by [{\it cf.}\ Eqs.(\ref{g1_pn}) and 
(\ref{a0_pn})]
\beq 
\int_0^1 dx \, G^p_1 (x, Q^2 ) \;\; = \;\; \frac{1}{12}g_A^{(3)}  
+\frac{1}{36}g_A^{(8)} + \frac{1}{9} \; g_A^{(0)}  
\; +\;  O\left(\frac{1}{Q^2}\right) .   
\label{proton_spin} 
\eeq   
When the charmed quark is included, its contribution can be expressed
as a modification of the $SU(3)$ flavor--singlet axial coupling constant
in (\ref{proton_spin}):
\beq 
g_A^{(0)} \;\; = \;\; g_{A, {\rm light}}^{(0)} \; + \; 2 g_A^{(c)} ,   
\eeq 
where $g_{A, {\rm light}}^{(0)}$ contains the $SU(3)$--singlet
contribution from the light flavors only, and $g_A^{(c)}$
is defined as
\beq
\langle p, s |\bar c \gamma_\beta\gamma_5 c |p, s \rangle  
\;\; = \;\; -2 g^{(c)}_A s_\beta . 
\eeq
In the proton, where the charmed quarks occur in the form of virtual 
$\bar c c$ pairs, the charm quark axial current operator can be 
approximated by an operator involving only light quarks ($u, d, s$)
and the gluon fields, using the heavy--quark mass ($1/m_c$--) expansion
to integrate out the charm degrees of freedom.
Refs.\cite{Franz:1999hw,Polyakov:1999rb}, correcting earlier results of
Refs.\cite{Halperin:1997jq,Araki:1998yk}, quote the following
result for the nucleon matrix element of the charm axial current
to order $1/m_c^2$:
\beq  
\langle p, s |\bar c \gamma_\beta \gamma_5 c |p, s \rangle  
\;\; = \;\; \frac{\alpha_s}{12\pi m_c^2} \; \langle p, s |
\sum_{f = u, d, s} \bar{\psi}_f  
\gamma_\alpha \Fdual_{\beta\alpha} \psi_f |p, s \rangle ,
\label{charm}  
\eeq   
from which follows that
\beq  
g_A^{(c)} \;\; =\;\; 
-\frac{\alpha_s}{12\pi}\frac{M_N^2}{m_c^2} f_2^{(0)}.  
\label{axial_charm}  
\eeq  
Using our result for the $SU(3)$ flavor--singlet matrix element,
$f_2^{(0)}$, Eq.(\ref{f20_su3}), we can now estimate the 
charm axial constant, $g_A^{(c)}$. With $\alpha_s (\mu = m_c ) = 0.39$
and $m_c = 1.15 \ldots 1.35 \, {\rm GeV}$, we find
\beq  
g_A^{(c)} \simeq -1\times 10^{-4}.  
\eeq     
The charm quark contribution to the proton spin is very small,
mostly because of the smallness of the flavor--singlet twist--4 matrix
element, $f_2^{(0)}$.
\section{Summary}  
The instanton vacuum with its inherent small parameter --- the packing fraction
of the instantons, $\bar\rho / \bar R$ --- implies a parametrical (and  
numerical) hierarchy of the spin--dependent twist--3 and 4 matrix elements:  
$\mbox{twist--3} \ll \mbox{twist--4}$. The unusually small value for the 
twist--3 matrix element $d^{(2)} \sim 10^{-3}$ obtained from the 
instanton vacuum \cite{Balla:1998hf} agrees well with the precise 
measurements of $G_2$ in the E155x \cite{Bosted:2000tf} experiment 
(and, also, the older E143 \cite{Abe:1998wq} and E155 \cite{Anthony:2000fn} 
results). Note also that recently revised lattice calculations of
$d_2$ \cite{Gockeler:2001ja}, properly accounting for operator mixing 
effects, confirmed the small value predicted by the instanton vacuum.
\par
In this paper we have presented numerical estimates of the 
twist--4 matrix elements $f_2$, including also its flavor--singlet
component. Our approach predicts a big numerical difference between the 
flavor--nonsinglet and singlet matrix elements, $|f_2^{(3)}| \gg |f_2^{(0)}|$,
in contrast to QCD sum rules, which tend to give values of the same
order of magnitude. Phenomenologically, this would imply much larger power 
corrections to the Bjorken than to the Ellis--Jaffe sum rule, 
a prediction which can in principle be tested experimentally.
We have also pointed out that the small value of $f_2^{(0)}$ implies
a very small intrinsic charm contribution to the nucleon spin.
\par
The approach laid out here can be extended in many  
ways; for example, one can compute also the matrix elements determining  
power corrections to higher moments of the structure functions and  
restore the the $x$--dependence of the twist--4 contribution.  
This would allow a more direct comparison of the higher--twist  
corrections with experimental data.  
\end{document}